\documentclass[aps,prb,twocolumn, superscriptaddress]{revtex4-2}

\usepackage[T1]{fontenc}
\usepackage[latin9]{inputenc}
\usepackage{color}
\usepackage{verbatim}
\usepackage{amsmath}
\usepackage{amssymb}
\usepackage{graphicx}

\usepackage{bm}

\usepackage{tikz}
\usepackage{diagbox}
\usepackage[normalem]{ulem}
\usepackage{wasysym}

\makeatletter
\PassOptionsToPackage{caption=false}{subfig} 
\usepackage{hyperref}
\hypersetup{
breaklinks=true,
colorlinks=true,
citecolor=blue,
linkcolor=black,
filecolor=black,
urlcolor=blue
}
\IfFileExists{lmodern.sty}{\usepackage{lmodern}}{}

\renewcommand{\bar}{\overline}
\renewcommand{\tilde}{\widetilde}

\renewcommand{\Im}{\operatorname{Im}}

\newcommand{\SU}{\operatorname{SU}}

\newcommand{\ZZ}{\mathbb{Z}}

\newcommand{\calH}{\mathcal{H}}

\newcommand{\calM}{\mathcal{M}}

\newcommand{\dd}{\mathrm{d}}

\makeatletter

\newcommand*{\wideboxed}[1]{\setlength{\fboxsep}{1ex}%
  \fbox{\m@th$\displaystyle#1$}}
\makeatother

\def\be{\begin{equation}}
\def\ee{\end{equation}}

\makeatother

\begin{document}

\title{
Space of conformal boundary conditions from the view of higher Berry phase:\\
Flow of Berry curvature in parametrized BCFTs\\
}

\author{Xueda Wen}
\affiliation{School of Physics, Georgia Institute of Technology, Atlanta, GA 30332, USA}

\date{Jul 16, 2025}

\begin{abstract}

In this work, we study the connection between two subjects: the space of conformal boundary conditions in boundary conformal field theories (BCFTs) and the space of gapped systems characterized by higher Berry phases. We explore this connection by analyzing multi-parameter spectral flow in Dirac fermion BCFTs with continuously parametrized conformal boundary conditions, which are introduced by coupling a CFT to a family of gapped systems. When the gapped systems belong to a nontrivial higher Berry class, the associated conformal boundary conditions induce a flow of the ordinary Berry curvature, resulting in a Chern number pump in the Fock space of the BCFT. This phenomenon is the BCFT analog of Berry curvature flow in one-dimensional parametrized gapped systems, where the flow occurs in real space. Building on this correspondence, we introduce the notions of higher Berry curvature and higher Berry invariants within the BCFT framework.
Our results provide a new perspective for studying the topological properties of families of conformal boundary states and gapped ground states: if a family of gapped states belongs to a nontrivial higher Berry class, then the corresponding entanglement Hamiltonians exhibit a multi-parameter spectral flow that carries Berry curvature in the Fock space.

\end{abstract}

\maketitle

\begin{center}
\textbf{Introduction}
\end{center}

Boundary conformal field theories (BCFTs) have been extensively studied since the 1980s. When the bulk CFT is rational with respect to a given chiral algebra, the set of conformal boundary conditions, or equivalently conformal boundary states, is typically discrete \cite{Cardy_1989}. In contrast, more general (non-rational) CFTs often admit a continuous family of conformal boundary conditions. 
The structure of this space has been explored in depth in \cite{Friedan_1993,Friedan_2003,1994_Callan,1998_Recknagel,2001_Gaberdiel,Gaberdiel_2002,Book_BCFT_2013}, revealing rich and intricate connections to boundary critical phenomena and D-brane dynamics.

Meanwhile, one forefront of modern condensed matter physics is to understand the space of gapped many-body systems
\cite{
kitaevSimonsCenter1,
kitaevSimonsCenter2,
kitaev2015talk,kitaev2019,
KS2020_higherberry, KS2020_higherthouless, Hsin_2020, Cordova_2020_i, Cordova_2020_ii, Else_2021,2022aBachmann,Choi_Ohmori_2022, 2023_Wen,Aasen_2022, 
	Hsin_2023,Kapustin2201, Shiozaki_2022, Ohyama_2022, ohyama2023discrete, Kapustin2305, homotopical2023, 2023Ryu, 2023_Qi,2023Shiozaki,2023Spodyneiko,2023Debray,
2024_Sommer1,2024_Sommer2,2024_Shuhei1,2024_Shuhei2,
2024_Multi_WF,2024_Geiko,
beaudry_2025,Kapustin_2025,Manjunath_2025,bose2025,2025_Jones,kubota2025}.
This interest has been largely driven by Kitaev's conjecture \cite{kitaevSimonsCenter1,
kitaevSimonsCenter2,
kitaev2015talk}, which proposes that invertible phases of matter are classified by generalized cohomology theories.
To probe the topology of the space of gapped 
many-body systems, Kapustin and Spodyneiko proposed the concept of higher Berry curvatures \cite{KS2020_higherberry,KS2020_higherthouless}, 
which was studied systematically in the framework of operator algebras for general cases with and without symmetries \cite{Kapustin2201}. Physically, the higher Berry curvature in $d$-dimensional systems can be interpreted as the flow of 
higher Berry curvature in $(d-1)$-dimensional systems \cite{2023_Wen}.
In one-dimensional systems, this reduces to a flow of ordinary Berry curvature in real space, giving rise to the intriguing phenomenon of Chern number pumping \cite{2023_Wen,2024_Sommer1}. Recent theoretical developments have deepened our understanding of higher Berry curvatures, from both Hamiltonian and wavefunction perspectives
\cite{kitaev2019,
KS2020_higherberry, KS2020_higherthouless, Hsin_2020,  Choi_Ohmori_2022, 2023_Wen,Aasen_2022, 
	Hsin_2023,Kapustin2201, Shiozaki_2022, Ohyama_2022, ohyama2023discrete, Kapustin2305, homotopical2023, 2023Ryu, 2023_Qi, 2023Shiozaki,2023Spodyneiko,2023Debray,
2024_Sommer1,2024_Sommer2,2024_Shuhei1,2024_Shuhei2,
2024_Multi_WF,2024_Geiko,
beaudry_2025,Kapustin_2025}.

\smallskip
In the introduction above, we identified two distinct ``spaces'': the space of conformal boundary conditions in BCFT and the space of gapped systems. Although these spaces originate from different physical systems, i.e., the gapless versus the gapped systems, it is interesting to ask whether there exists any connection between them. In particular, if such a connection does exist, can topological properties in parametrized families of gapped systems, such as the Berry curvature flow \cite{2023_Wen}, manifest themselves within the framework of BCFT?
The main goal of this work is to address these questions by investigating the relation between parameterized BCFTs and parameterized gapped systems. While our main focus is on (1+1)-dimensional spacetime, the framework we develop extends naturally to higher dimensions.

\begin{figure}
\centering
\includegraphics[width=2.2in]{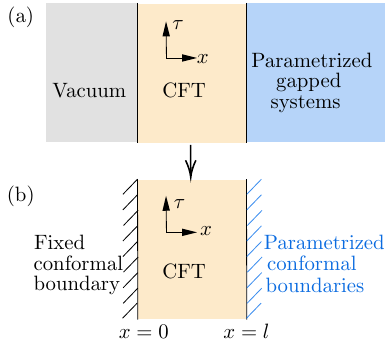}
\caption{(a) A CFT defined in the interval $[0,l]$ sandwiched by a vacuum and a parametrized family of gapped systems, which have the \textit{same} spacetime dimensions as the CFT.
(b) As the energy gaps in the vacuum and gapped systems go to infinity, we have a BCFT with a fixed conformal boundary condition on the left and a parametrized family of conformal boundary conditions on the right.}
\label{Fig:BCFT}
\end{figure}

Our setup is illustrated in Fig.\ref{Fig:BCFT}. We begin by introducing a parametrized family of gapped systems. Instead of terminating these gapped systems directly at the vacuum, we insert a CFT between the vacuum and the gapped region. As the parameters of the gapped systems are varied, a topological quantum pump may occur if the systems belong to a nontrivial higher Berry class. In the case of the $(1+1)d$ systems studied in \cite{2023_Wen}, the ordinary Berry curvature can flow from the bulk to the boundary, and this flow is expected to produce observable signatures within the CFT region.

To study the physics described above more transparently, we consider a parametrized family of BCFTs as illustrated in Fig.~\ref{Fig:BCFT}(b). The boundary condition on the left is fixed, while on the right, we have a continuous family of conformal boundary conditions, which can be obtained from the setup in Fig.~1(a) by taking the energy gap of the gapped systems to infinity. As we will show, this configuration allows the topological features of the gapped systems to manifest through the multi-parameter spectral flow \cite{2023spectral} in the BCFT. In particular, the Berry curvature flow and the resulting Chern number pump can be observed in parametrized BCFTs, where the flow occurs not in real space, but in the Fock space. This result will have direct applications in understanding the topological properties of 
families of \textit{gapped} states.

We emphasize that the sandwich structure in Fig.~\ref{Fig:BCFT} differs from those commonly studied in the context of symmetry topological field theory, where a $d$-dimensional quantum system is sandwiched between two $(d+1)$
dimensional topological field theories. See, e.g., \cite{Ji_Wen_2020,2209_Freed,2023_Huang_Cheng}. In contrast, in our setup, all layers reside in the same spacetime dimension.

\begin{center}
\textbf{Warm up: $U(1)$ Thouless pump and
parametrized BCFTs}
\end{center}

As a warm up, let us start with $U(1)$ charge pump in parametrized BCFTs. 
We consider a parametrized family of gapped systems with $U(1)$ Thouless charge pump in Fig.\ref{Fig:BCFT} (a), with the Hamiltonian density \cite{Hsin_2020}
\be
\label{H_U1}
\calH_{U(1)}=\Psi^\dag(x)(-i\sigma_3 \partial_x+m_1 \sigma_1+m_2\sigma_2)\Psi(x),
\ee
where 
$\Psi(x)=(\psi_R(x),\,\psi_L(x))^T$ is the Dirac fermion operator with right and left moving components, and $\sigma_i$ are the Pauli matrices.
We choose the masses $m_1$ and $m_2$ in the parameter space $X=S^1$ as 
\be
\label{X_S1}
m_2+im_1=:m \, e^{i\alpha}, \quad \alpha\in[0,2\pi],
\ee
where $m>0$ is fixed. As we adiabatically change $\alpha$ from 0 to $2\pi$, a quantized charge will be pumped along this $1d$ system, which is known as Thouless pump \cite{thouless_1983}.

Now we connect the gapped systems with a CFT,
as shown in Fig.\ref{Fig:BCFT}.
The CFT Hamiltonian density $\calH_{\text{CFT}}$ is obtained from \eqref{H_U1} by turning off the mass term. This CFT has a central charge $c=1$ with $U(1)\times U(1)$ symmetry, which arises from the conservation of 
charges for the chiral and anti-chiral sectors, respectively.
By introducing a boundary which couples the chiral and anti-chiral modes, at least one $U(1)$ symmetry will be broken. 

In Eq.\eqref{X_S1}, if we take the infinite mass limit, i.e., $m\to \infty$, then the gapped systems in \eqref{H_U1} will result in a continuous one-parameter family of conformal boundary conditions at $x=l$ \cite{Appendix}:
\be
\label{U1_boundary}
\psi_R(x)=e^{i\alpha} \psi_L(x),\quad \alpha\in[0,2\pi],
\ee
where $\alpha\in X=S^1$ is the parameter in \eqref{X_S1}.  
Then at the other end $x=0$, we impose a fixed boundary condition with $\psi_L(x)=\psi_R(x)$.

Now the problem is reduced to the study of a BCFT with conformal boundary conditions in Fig.\ref{Fig:BCFT} (b). With conformal boundaries, only a single $U(1)$ symmetry remains, corresponding to total fermion number conservation.
The fermion fields satisfying the above boundary conditions can be mode expanded as
\be
\label{Mode_expansion}
\small
\psi_{R,L}(x)=\sqrt{\frac{\pi}{l} }\sum_{r\in \mathbb Z+\alpha/2\pi}\psi_r\, e^{\pm i\frac{\pi}{l} r x},
\ee
where $\{\psi_{r},\psi^\dag_{r'}\}=\delta_{r,r'}$ and 
$\{\psi_{r},\psi_{r'}\}=\{\psi^\dag_{r},\psi^\dag_{r'}\}=0$.
The left and right moving fields are expanded with the same modes $\psi_r$, since they are coupled to each other at the two boundaries. Then the CFT Hamiltonian has the form
\be
\label{H_U1_diagonal}
\small
H_{\text{CFT}}
=\int_0^l \calH_{\text{CFT}} (x)\,\dd x
=
\frac{\pi}{l} \sum_{r\in \mathbb Z+\alpha/2\pi} r\, \psi_r^\dag \psi_r.
\ee
The family of ground states of \eqref{H_U1_diagonal} depend on $\alpha$ as
\be
\label{G_alpha}
|G\rangle_\alpha=\prod_{r\le 0} \psi_r^\dag |\text{vac}\rangle, \quad r\in \mathbb Z+\alpha/2\pi.
\ee
As we adiabatically increase $\alpha$ from 0 to $2\pi$, both the energy spectrum and the fermion number evolve continuously -- this process is known as a one-parameter \textit{spectral flow}. At $\alpha=2\pi$, the energy spectrum returns to its original form at $\alpha=0$, but the total fermion number increases by one, manifesting a nontrivial spectral flow.

Based on \eqref{H_U1_diagonal} and \eqref{G_alpha}, careful readers may have noted that the BCFT setup can be mapped to a chiral CFT on a circle with twisted boundary conditions \cite{hori2003mirror,Appendix}. The underlying physics, however, is quite different.
In the chiral CFT case, the pumped charge  originates from a $(2+1)d$ gapped system with a nonzero integer Hall conductivity. In contrast, in our parametrized BCFTs, the pumped charge arises from $(1+1)d$ gapped systems that share the same spacetime dimension as the BCFTs.

\begin{center}
\textbf{Berry curvature flow and Chern number pump\\
in parametrized BCFTs}
\end{center}

Let us now turn to the central example of this work: BCFTs with a parametrized family of conformal boundary conditions exhibiting Berry curvature flow and Chern number pumping.

\begin{figure}[tp]
\centering
\includegraphics[width=2.8in]{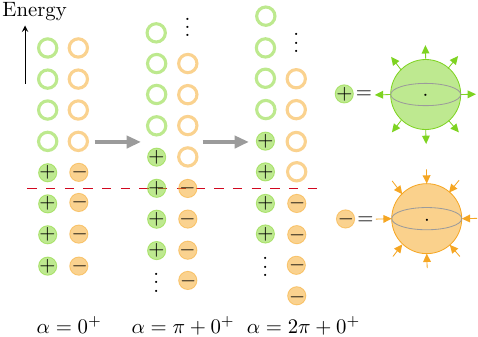}
\caption{Berry curvature flow and Chern number pump in parametrized BCFTs.
The energy spectrum can be divided into two groups based on the signs of their Chern numbers, with each occupied mode $\oplus$/$\ominus$  carries a ``$+1$''/``$-1$'' Chern number.
Each $\ocircle$ denotes a mode that is not occupied by a fermion.
As we tune the three parameters $(\alpha,\, \theta,\, \phi)$, 
the energy spectrum only depends on $\alpha$, while the 2-form Berry curvature for each occupied mode is only determined by $\theta$ and $\phi$ as 
$\Omega^{(2)}_{\pm}=\pm \frac{\sin\theta}{2}\dd\theta\wedge \dd\phi$, the integral of which over $X=S^2$ spanned 
by $\theta$ and $\phi$ (right plot) gives a quantized 
Chern number $\pm 1$.
When $\alpha$ is increased by $2\pi$, the energy spectrum and the total fermion number remain unchanged, but the total Chern number shifts by two.
}
\label{Fig:Chern-Pump}
\end{figure}

In Fig.\ref{Fig:BCFT}, 
we consider the family of gapped systems that belong to a nontrivial higher Berry class, with Hamiltonian density given by
$
\calH_{\text{HB}}=\Psi^\dag(x) H \Psi(x)
$
with \cite{2000_Abanov_Wiegmann,Hsin_2020}
\be
\label{H_HB}
H=-i(\sigma_0\otimes \sigma_3)\partial_x+\Big(m_0\,  \sigma_0\otimes \sigma_2+\sum_{k=1}^3 m_k \,  \sigma_k\otimes \sigma_1  \Big),
\ee
where $\Psi=(\psi_{1,R},\, \psi_{1,L}, \,\psi_{2,R}, \, \psi_{2,L})^T$\footnote{
Our convention of $H$ is slightly different from those in 
Ref.\cite{2000_Abanov_Wiegmann,Hsin_2020}.
}. 
The parameter space is chosen as $X=S^3$, such that 
$\sum_{i=0}^3 m_i^2=m^2$ is fixed.
This model, which was studied two decades ago \cite{2000_Abanov_Wiegmann}, was recently discussed in the 
framework of higher Berry phase in Ref.\cite{Hsin_2020}.
Here we will show that the conformal boundary conditions 
introduced by this family of gapped systems will result in Berry curvature flow in the corresponding BCFTs.

For later convenience, we parametrize the mass terms in Eq.\eqref{H_HB} using hyperspherical coordinates on the three-sphere $X=S^3$:
\be
\label{HyperSphericalCoordinateM}
\begin{split}
&m_0/m=\cos\alpha,\quad
m_1/m=\sin\alpha\sin\theta\cos\phi\\
&m_2/m=\sin\alpha\sin\theta\sin\phi,\,\,\,
 m_3/m=\sin\alpha \cos\theta,
\end{split}
\ee
where $\alpha,\,\theta\in[0,\pi]$, and $\phi\in [0,2\pi]$.
In the CFT region, the Hamiltonian density is obtained by turning off the mass term in \eqref{H_HB}:
\be
\label{H_CFT_Chern}
\mathcal H_{\text{CFT}}=\sum_{a=1}^2 \big[-i\psi_{a,R}^\dag(x) \partial_x \psi_{a,R}(x)
+i \psi_{a,L}^\dag(x) \partial_x \psi_{a,L}(x)\big].
\ee
That is, we have two copies of Dirac fermion CFTs and the total central charge is $c=2$.
For the configuration in Fig.\ref{Fig:BCFT},
by taking the infinite mass limit in \eqref{H_HB},
one can find the conformal boundary conditions at $x=l$ become:
\be
\label{S3_boundary}
\Psi_R(x)= M\cdot \Psi_L(x),
\ee
where $M$ is an $\SU(2)$ matrix \cite{Appendix}
\be
\label{S3_boundary2}
M=\begin{pmatrix}
\cos\alpha+i\sin\alpha\cos\theta &i e^{i\phi} \sin\alpha \sin\theta\\
i e^{-i\phi} \sin\alpha \sin\theta &\cos\alpha-i\sin\alpha\cos\theta
\end{pmatrix}.
\ee
That is, this family of conformal boundary conditions are parametrized by $(\alpha,\theta,\phi)\in X=S^3$. 
At the left boundary $x=0$, we impose the 
fixed conformal boundary condition $\Psi_R(x)=\Psi_L(x)$.
Now let us consider the following unitary transformation
\be
\label{U_transform}
\Psi_{R,L}(x)=U\, \tilde\Psi_{R,L}(x),
\ee
where $U$ \textit{only} depends on $\theta$ and $\phi$ as follows
\be
\small
\label{U_matrix}
U(\theta,\phi)=\begin{pmatrix}
\cos(\theta/2) &-e^{i\phi} \sin(\theta/2) \\
e^{-i\phi}\sin(\theta/2) &\cos(\theta/2) 
\end{pmatrix}
\ee
such that the twisted boundary conditions in \eqref{S3_boundary} become diagonal:
\be
\label{S3_bdry_diagonal}
\tilde{\psi}_{1,R}(x)=e^{i\alpha}\tilde{\psi}_{1,L}(x), \quad 
\tilde{\psi}_{2,R}(x)=e^{-i\alpha}\tilde{\psi}_{2,L}(x),
\ee
 while the boundary condition at $x=0$ is still fixed as $\tilde\Psi_R(x)=\tilde\Psi_L(x)$.
By considering mode expansions similar to \eqref{Mode_expansion} \cite{Appendix}, 
the total Hamiltonian corresponding to \eqref{H_CFT_Chern} becomes
\be
\label{H_Chern_mode}
H_{\text{CFT}}= \frac{\pi}{l} \sum_{r\in \mathbb Z+\alpha/2\pi} r\, \tilde \psi_{1,r}^\dag \tilde\psi_{1,r}
+
\frac{\pi}{l}
\sum_{s\in \mathbb Z-\alpha/2\pi} s \, \tilde \psi_{2,s}^\dag \tilde\psi_{2,s},
\ee
where the fermionic fields satisfy
$\{\tilde\psi_{1,r},\tilde\psi^\dag_{1,r'}\}=\delta_{r,r'}$, 
and 
$\{\tilde\psi_{1,r},\tilde\psi_{1,r'}\}=\{\tilde\psi^\dag_{1,r},\tilde\psi^\dag_{1,r'}\}=0$, and similarly for $\tilde\psi_{2,s}$ and $\tilde\psi^\dag_{2,s}$.
Comparing \eqref{H_Chern_mode} with \eqref{H_U1_diagonal}, 
one might naively interpret the system as comprising two copies of CFTs, each undergoing a single-parameter spectral flow in opposite directions.
However, one should be careful here, because the ground state 
of $H_{\text{CFT}}$ in \eqref{H_Chern_mode} depends on three parameters rather than just one. More explicitly, the ground state of $H_{\text{CFT}}$ in \eqref{H_Chern_mode} is
\be
\label{G_Chern}
|G\rangle_{\alpha,\theta,\phi}=
\prod_{r\le 0} \tilde\psi_{1,r}^\dag 
\prod_{s\le 0} \tilde\psi_{2,s}^\dag |\, \text{vac}\rangle,
\ee
where $r\in\ZZ+\alpha/2\pi$ and $s\in\ZZ-\alpha/2\pi$. 
To explicitly see the parameter dependence,  let us examine closely the single-mode state  $\tilde \psi_{1,r}^\dag |\text{vac}\rangle$, which takes the form:
\be
\label{WF:mode}
\small
\frac{1}{\sqrt{4\pi l}} \sum_{a=1,2}U_{a1}
\int_0^l
\dd x
\left(
\psi^\dag_{a,R}(x) e^{+i\frac{\pi r}{l} x}+\psi^\dag_{a,L}(x) e^{-i\frac{\pi r}{l}  x}
\right)
 |\text{vac}\rangle,
\ee
where $\psi_{a,R}(x)$ and $\psi_{a,L}(x)$ are the original fermion fields in \eqref{H_CFT_Chern}, 
$U_{a1}$ denotes the first column in \eqref{U_matrix} and depends on both $\theta$ and $\phi$,
while the dependence on $\alpha$ enters through the mode 
index $r\in\ZZ+\alpha/2\pi$.
For this single-mode wavefunction,
one can show that it carries a nontrivial two-form Berry curvature given by:
\be\label{Omega2}
\Omega_+^{(2)}=\Omega_{\theta\phi}\, \dd \theta\wedge \dd \phi=
\frac{\sin\theta}{2}\dd\theta\wedge \dd\phi.
\ee
That is, only the component $\Omega_{\theta\phi}$ is nonzero, while the other two components, $\Omega_{\alpha\theta}$ and $\Omega_{\alpha\phi}$, vanish. Similarly, for the single-mode $\tilde{\psi}^\dag_{2,s}|\text{vac}\rangle$ in \eqref{G_Chern}, one can verify that it carries a two-form Berry curvature 
$\Omega_-^{(2)}=-\frac{\sin\theta}{2}\dd\theta\wedge \dd\phi$, 
which differs from Eq.~\eqref{Omega2} only by an overall minus sign. $\Omega_{\pm}^{(2)}$ give rise to quantized Chern numbers associated with the single-particle modes as 
\be
\label{Chern_number}
\text{Ch}_{\pm}=\frac{1}{2\pi}\int_{S^2}\Omega_\pm^{(2)}=\pm 1,
\ee
where the integral is taken over the two-sphere $S^2$ 
parametrized by $\theta$ and $\phi$, as illustrated in Fig.\ref{Fig:Chern-Pump}.
A subtle point arises at $\alpha=0$ and $\pi$ on the manifold  $X=S^3$, where the two-sphere $S^2$ collapses to a point, rendering $\Omega_{\pm}^{(2)}$ ill-defined. Physically, 
 this occurs because the twisted boundary conditions in Eq.~\eqref{S3_boundary2} reduce to the identity (up to a minus sign), becoming independent of $\theta$ and $\phi$. One can consider an arbitrary unitary transformation in \eqref{U_transform} to diagonalize the Hamiltonian, and the associated Berry curvature at these points is not uniquely defined. This subtlety will play an important role when we define the higher Berry curvature later.

\smallskip
Now we are ready to consider the Berry curvature flow and Chern number pump in the Fock space of BCFT.
As illustrated in Fig.\ref{Fig:Chern-Pump}, the energy spectrum can be grouped into two sets of modes according to the sign of the Chern numbers they carry. We begin at
$\alpha=0^+$, where the ground state in Eq.\eqref{G_Chern} corresponds to a filled Fermi sea. As we vary the three parameters $(\alpha, \theta, \phi)$, the energy levels flow across zero. Specifically, $(\theta,\phi)$ determine the 2-form Berry curvature associated with each mode, while $\alpha$
controls the energy spectrum, as seen in Eq.~\eqref{H_Chern_mode}. When we adiabatically increase
$\alpha$ by $2\pi$ (but not $\pi$), and simultaneously let $(\theta,\phi)$ wrap around the two-sphere $S^2$, we find that both the energy spectrum and the total fermion number return to their original values. However, the total Chern number shifts by two, reflecting the fact that this process effectively wraps the full parameter space $X=S^3$ twice.
Physically, the increase in the total Chern number in this BCFT originates from the Chern number pump contributed by the parametrized gapped systems shown in Fig.\ref{Fig:BCFT},
where the Chern number is pumped from the bulk to the boundary \cite{2023_Wen} \footnote{In Ref.\cite{2023_Wen}, the term ``Chern number pump'' is used for a closely related system over $X=S^2\times S^1$, but the essential physics of both systems is the same.  Here we use the term ``pumping''
for $X=S^3$ somewhat more loosely.}.

\smallskip
A natural question arises: how can we characterize the flow of Berry curvature associated with each parameter in  $X=S^3$? To address this, we need to quantify how much Berry curvature is carried by the ground state. Since there are only two types of occupied fermion modes, each carrying either $\Omega^{(2)}_+$ or $\Omega^{(2)}_-$, it suffices to count the corresponding fermion numbers in the ground state, denoted by $\langle Q_+\rangle$ and $\langle Q_-\rangle$, respectively.
For example, let us consider $Q_+=\frac{1}{2\pi}\int_0^l (\tilde{\psi}_{1,R}^\dag \tilde{\psi}_{1,R}+\tilde{\psi}_{1,L}^\dag \tilde{\psi}_{1,L}) \dd x$, which can be further written as
\be
\label{Q+}
Q_+=\sum_{r\in \mathbb Z+\alpha/2\pi}: \tilde{\psi}_{1,r}^\dag \tilde{\psi}_{1,r}:
+\frac{\alpha}{2\pi}-\Big[\frac{\alpha}{2\pi}\Big]-\frac{1}{2}.
\ee
Here we have considered the zeta function regularization to sum up the ground state charges by using $\sum_{r\in \ZZ+a, r<0}=\zeta(0,1-(a-[a]))=a-[a]-1/2$, where $[a]$ is the greatest integer less than or equal to $a$, and $:\,:$ represents the normal ordering.
The total fermions $Q_-$ with each fermion carrying $\Omega^{(2)}_-$ can be written down similarly \cite{Appendix}.

Then the total 2-form Berry curvature contributed 
by the filled fermi sea in the ground states in \eqref{G_Chern} is 
\be
\small
\label{F2}
\omega^{(2)}=\langle Q_+\rangle\,\Omega_+^{(2)}+
\langle Q_-\rangle\, \Omega_-^{(2)}
=\left(\frac{\alpha}{\pi}-1\right)  \frac{\sin\theta}{2} \dd\theta\wedge \dd\phi,
\ee
based on which one can obtain the closed 3-form higher Berry curvature as
\be
\label{Omega3}
\Omega^{(3)}:=\dd \omega^{(2)}=\frac{1}{2\pi}\sin\theta\, \dd\alpha\wedge \dd\theta\wedge \dd\phi.
\ee
By choosing any reference energy in the 
fermion sea, e.g., the red dashed line in 
Fig.\ref{Fig:Chern-Pump}, the flow of Berry curvature acrosss this reference energy is captured by $\Omega^{(3)}$.  Based on Eq.\eqref{Omega3}, the associated quantized higher Berry invariant can be obtained by integrating the higher Berry curvature over the three-sphere $X = S^3$ as
\be
\int_{S^3} \Omega^{(3)} = 2\pi.
\ee

As a remark, from Eq.\eqref{Omega3}, one might think that $\Omega^{(3)}$ is an exact form rather than a closed one. However, $\Omega^{(3)}$ is indeed a closed form, since the two-form $\omega^{(2)}$ in Eqs.\eqref{F2} and \eqref{Omega3} is not globally well-defined over the three-sphere $X = S^3$. As discussed below Eq.\eqref{Chern_number}, $\omega^{(2)}$ becomes ill-defined at $\alpha = 0$ and $\alpha = \pi$. Taking this into account, one can also evaluate the higher Berry invariant using Stokes' theorem:
\be
\label{Stokes}
\small
\int_{\calM^3} \Omega^{(3)}=\int_{S^2_{\alpha=\pi-0^+}}\omega^{(2)}
-
\int_{S^2_{\alpha=0^+}}\omega^{(2)}=2\pi,
\ee
where $\calM^3$ is the three manifold obtained from $S^3$ by removing $\alpha=0$ and $\alpha=\pi$, and $S^2_{\alpha}$ represents a two-sphere $S^2 \subset S^3 $ with a fixed $\alpha$. From Eq.\eqref{Stokes}, it is evident that the higher Berry invariant quantifies the Chern number pump during the adiabatic process.

Finally, we highlight the structural similarity between Eqs.\eqref{F2} and \eqref{Omega3} and the real-space Berry curvature flow in Ref.\cite{2023_Wen}. In Ref.\cite{2023_Wen}, one can define $\omega^{(2)}$, which captures the total 2-form curvature within a boundary region in real space, with its exterior derivative $\dd\,\omega^{(2)}$ describing the Berry curvature flow in real space. Analogously, here the  $\omega^{(2)}$ defined in Eq.\eqref{F2} represents the total 2-form curvature within a ``boundary'' region in Fock space, i.e., all occupied states below a chosen reference energy, and $\dd\, \omega^{(2)}$ in Eq.\eqref{Omega3} characterizes the flow of 2-form ordinary Berry curvature in the Fock space.

The same idea above can be extended to compact free boson BCFTs with a general compactification radius that encodes
the strength of an effective four-fermion interaction.
The details of this generalization will be presented elsewhere.

\begin{center}
\textbf{Application in parametrized gapped states}
\end{center}

\begin{figure}[tp]
\centering
\includegraphics[width=3.2in]{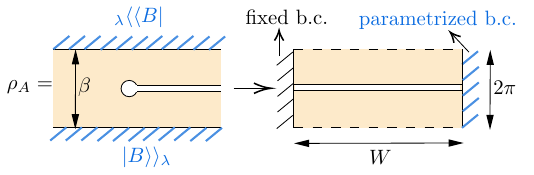}
\caption{Left: Reduced density matrix $\rho_A$ (in Euclidean spacetime $z=z+i\tau$) for region $A=[0,+\infty)$ in a family of regularized conformal boundary states in \eqref{B_lambda}.
A small disk of radius $\epsilon_0$ is removed at the entanglement cut $z=0$ to introduce a UV cut-off.
The fields living on the upper and lower edges of $[0,+\infty)$ correspond to the rows and columns of $\rho_A$.
Right: Under a conformal mapping $w=f(z)$, $\rho_A$ is mapped to a cylinder of circumference $2\pi$, 
with a fixed boundary condition at one end and 
parametrized boundary conditions at the other.
}
\label{Fig:rho}
\end{figure}

As an application of our results, we show how to extract
the topological properties of a family of conformal boundary states
$|B\rangle\rangle_\lambda$, parametrized by $\lambda\in X$.

Since each conformal boundary state $|B\rangle\rangle_\lambda$ is non-normalizable and exhibits zero real-space entanglement \cite{miyaji_2014}, 
we consider the regularized state
\be
\label{B_lambda}
|\psi\rangle_\lambda =e^{-\frac{\beta}{2} H_{\text{CFT}}}\, |B\rangle\rangle_\lambda, \quad \lambda\in X,
\ee
where $\beta>0$ is a regularization parameter controlling the correlation length of $|\psi\rangle_\lambda$, and $H_{\text{CFT}}$ denotes the CFT Hamiltonian.
The states in Eq.\eqref{B_lambda} can be interpreted as ground states of certain gapped Hamiltonians \cite{miyaji_2014,Cho_2017_A,Cho_2017,Cardy_2017,2017_Han,Hsieh_2024}, although a rigorous justification of this identification remains open. Physically, since gapped ground states may exhibit charge or Berry curvature flow in real space   \cite{2023_Qi,2024_Sommer1}, such flows are expected to manifest within a chosen subsystem 
$A=[0,+\infty)$. To probe this, we examine the parametrized family of reduced density matrices
\be
\rho_A(\lambda)=\text{Tr}_{\bar A}(|\psi\rangle_\lambda\, _\lambda\langle \psi|)=: e^{-2\pi H_E(\lambda)},
\ee 
where $H_E(\lambda)$ are the entanglement Hamiltonians. 
We consider a path-integral representation of $\rho_A$ in Euclidean space $z=x+i\tau$, as illustrated in Fig.~\ref{Fig:rho}. To introduce a UV cutoff, we remove a small disk of radius $\epsilon_0$ around the entanglement cut at $z=0$.
A fixed conformal boundary condition is imposed along the boundary of the removed disk \cite{Cardy_2016}.
Next, by applying the conformal map $w=f(z)=\log [\tanh(\pi z/2\beta)]$, $\rho_A$ is mapped to a cylinder of circumference $2\pi$ and width $W=\log [\coth(\pi\epsilon_0/2\beta)]$,  with two open ends: one end with a fixed boundary condition, and the other with parametrized boundary conditions. Then the entanglement Hamiltonian $H_E$ corresponds to the \textit{physical} Hamiltonian that generates translations along the vertical $\Im(w)$ direction, which is
precisely the same Hamiltonian discussed in Fig.\ref{Fig:BCFT} (b).
In other words, given a nontrivial parametrized family of conformal boundary states $|B\rangle\rangle_\lambda$, there will be a multi-parameter spectral flow in the corresponding parametrized entanglement Hamiltonians $H_E$.

We show in \cite{Appendix} that a similar result holds for families of gapped ground states near a critical point, derived via a complementary approach. These findings highlight the reduced density matrix, or equivalently the entanglement Hamiltonian, is a key object in the study of parametrized gapped systems, aligning with perspectives advocated in Refs.\cite{2023_Qi,2024_Sommer1}.
In particular, we expect that classifying the space of reduced density matrices for parametrized gapped 
systems may be closedly related to classifying the space of conformal boundary conditions/states. 

\begin{center}
\textbf{Conclusion}
\end{center}

To conclude, we have explicitly demonstrated the physics of Berry curvature flow and Chern number pumping arising from a continuous family of conformal boundary conditions in $(1+1)d$ Dirac fermion BCFTs. Physically, this multi-parameter spectral flow in the Fock space of BCFTs originates from the real-space Berry curvature flow in associated gapped systems. Our results can be applied to parametrized familied of gapped states: there will be a (multi-parameter) spectral flow in the entanglement Hamiltonians if the family of gapped states is in a nontrivial higher Berry class.

Given that the higher Berry curvature persists in interacting systems \cite{KS2020_higherberry}, it is natural to investigate its realization in more general CFTs. 
This will bring new insights into the rich structures uncovered in the space of conformal boundary states and conditions as explored in Refs.~\cite{1994_Callan,1998_Recknagel,2001_Gaberdiel,Gaberdiel_2002,Book_BCFT_2013}.
It is also natural to consider higher-dimensional BCFTs within the same framework, where the physics becomes even more intriguing as it may host rich phenomena such as higher Thouless pump \cite{KS2020_higherthouless}. 
Furthermore, for studying $(2+1)$-dimensional parameterized gapped systems, one may employ parameterized conformal boundary states or Ishibashi states in lower-dimensional, i.e., $(1+1)$-dimensional, CFTs, as suggested by the framework in Refs.~\cite{2012_Qi,2015_Das,Wen_2016}.

Another promising direction is to develop a practical method for detecting the higher Berry phase based on our setup. The setup illustrated in Fig.\ref{Fig:BCFT} can be viewed as a lead coupled to the gapped systems, with the conformal boundary conditions corresponding to the reflection matrix describing boundary scattering. Since the reflection matrix can, in principle, be probed through interference experiments, it may serve as an experimental handle for constructing the topological invariants of the gapped systems \cite{2025_Lo_Wen}.

\medskip
\textit{Note added:}
While completing this manuscript, we became aware of an upcoming related work Ref.\cite{2025_Choi_Ryu}, which will appear on arXiv on the same day.

\begin{center}
\textbf{Acknowledgments}
\end{center}

 The author thanks for discussions and related collaborations with Agnes Beaudry, Bo Han, Michael Hermele, Chih-Yu Lo, Juan Moreno, Markus Pflaum, Marvin Qi, Daniel Spiegel, David Stephen, Ophelia Sommer, and Ashvin Vishwanath, and thanks Roman Geiko and Yichen Hu for an interesting discussion on the connection between T-duality and higher Berry phase. 
 The author also thanks the authors of \cite{2025_Choi_Ryu} for coordinating submissions of our papers to arXiv. 
 This work is supported by a startup at Georgia Institute of Technology.

\bibliography{pump}

\appendix

\section{From boundary scattering to conformal boundary conditions}
\label{Appendix:BC}

In this appendix, we give details on how to determine the conformal boundary conditions in \eqref{U1_boundary} and \eqref{S3_boundary}.

\begin{figure}[tp]
\centering
\includegraphics[width=2.6in]{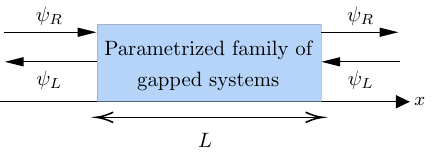}
\caption{A parametrized family of gapped systems in the region $[0,L]$ are connected to two CFTs in $(-\infty,0)$ and $(L,+\infty)$, respectively. 
Plane waves in the CFT regions are scattered by the gapped system. When the energy of an incoming wavefunction lies below the bulk energy gap, the wavefunction is completely reflected -- with reflection amplitude $|r|=1$ -- in the limit $L\to\infty$.
}\label{Fig:Scattering}
\end{figure}

We consider the scattering setup illustrated in Fig.\ref{Fig:Scattering}, where the parametrized gapped system is confined to the region $[0,L]$, and the CFTs are defined in the semi-infinite regions $(-\infty,0)$ and $(L,+\infty)$, respectively. In general, the wavefunctions on either side of the gapped region are related by a scattering matrix \cite{datta1997electronic}.
In this appendix, we focus on the limit $L\to \infty$. 
Then any incident wavefunction with energy below the bulk energy gap will be completely reflected at the interface.

We begin by considering the parametrized gapped systems described in Eq.\eqref{H_U1}. In this case, it suffices to consider a single copy of a complex fermion in the CFT region.
An eigenstate $\Psi$ of the full Hamiltonian with energy $E$ satisfies 
\be
\Psi(L)=\exp[(i E \sigma_3 +m_1\sigma_2-m_2\sigma_1)L] \Psi(0).
\ee
By substituting $\Psi(0)=(1,\,r )^T$ and $\Psi(L)=(t,\,0)^T$, one can obtain both the reflection coefficient $r$ and the transmission coefficient $t$. Here we are interested in the limit $L\to \infty$. 
When the energy $E$ is below the gap, i.e., $|E|<m$, 
the right-moving mode $\psi_R$ is completely reflected 
into the left-moving mode $\psi_L$ at $x=0$, with reflection amplitude satisfying $|r|=1$. More explicitly, one can find
\be
r=\frac{(m^2-E^2)^{1/2}+iE}{m}e^{-i\alpha},
\ee
where $|E|<m$ and $\alpha$ is defined in \eqref{U1_boundary}. Apparently, the phase of the reflection coefficient $r$ depends on the energy of the incoming wave, rendering the boundary condition 
non-scale-invariant.
To recover conformal boundary conditions, we take the limit
$m\to +\infty$, in which case the reflection coefficient becomes
\be
r=e^{-i\alpha},
\ee
or equivalently, the boundary condition at $x=0$ takes the form $\psi_L(x)=e^{-i\alpha} \psi_R(x)$. This yields the conformal boundary condition given in Eq.\eqref{U1_boundary}.

\smallskip

Now let us consider the parametrized gapped systems in \eqref{H_HB}. In this case, we need to introduce two copies of Dirac fermions in the CFT regions, and the reflection $r$ is described by a matrix rather than a number. One can use the same approach as introduced above to obtain $r$. In the limit $m\to +\infty$ and $L\to +\infty$, one can find that 
\be
\Psi_L(x=0)=r \, \Psi_R(x=0),
\ee
where 
\be
r= \frac{1}{m}
\begin{pmatrix}
m_0-im_3 &m_2-im_1\\
-m_2-im_1 &m_0+im_3
\end{pmatrix}
\in \SU(2).
\ee
Here the masses $m_i$ are those in the gapped Hamiltonians in \eqref{H_HB}, with $\sum_{i=0}^3 m_i^2=m^2$ where $m>0$.
Equivalently, we can write 
\be
\Psi_R(x=0)=r^{-1} \, \Psi_L(x=0),
\ee
which gives the expression in \eqref{S3_boundary} by considering the hyperspherical coordinates
in \eqref{HyperSphericalCoordinateM}.

\section{More details on Berry curvature flow in BCFTs}


For the Hamiltonian in \eqref{H_Chern_mode}, we have considered the mode expansion as
\be
\small
\tilde\psi_{1,r}=
\frac{1}{\sqrt{4\pi l}}\Big( \int_{0}^l \tilde\psi_{1,R}(x) e^{-i\frac{\pi}{l} rx}\dd x+ \int_{0}^l \tilde\psi_{1,L}(x) e^{i\frac{\pi}{l} rx}\dd x\Big),
\nonumber
\ee
and similarly for $\tilde\psi_{2,s}$.
Then based on the untiary transforamtion in \eqref{U_transform}, one can express $\tilde\psi_{1,r}$
and $\tilde\psi^\dag_{1,r}$ in terms of the original fermion fields in \eqref{WF:mode}, and similarly for $\tilde\psi_{2,s}$ and $\tilde\psi_{2,s}^\dag$.

Now let us consider the 2-form Berry curvature of the single-mode wavefunction in \eqref{WF:mode}, which we denote as $|u_{1,r}\rangle:=\tilde \psi_{1,r}^\dag |\text{vac}\rangle$. It carries a 2-form Berry curvature $\Omega^{(2)}_+=i\langle \dd u_{1,r}|\dd u_{1,r}\rangle$, where $\dd:=\sum_\lambda  \dd \lambda \frac{\partial}{\partial \lambda}$, with $\lambda\in (\alpha,\theta,\phi)$. After a straightforward calculation, one can find $\Omega_+^{(2)}$ in \eqref{Omega2}.

To study the total 2-form Berry curvature in the ground state, one needs to regularize the charges $Q_{+(-)}=\frac{1}{2\pi}\int_0^l (\tilde{\psi}_{1(2),R}^\dag \tilde{\psi}_{1(2),R}+\tilde{\psi}_{1(2),L}^\dag \tilde{\psi}_{1(2),L}) \dd x$. 
In terms of the Fourier modes $\tilde\psi_{1,r}$ and $\tilde\psi_{2,s}$, 
one can find $Q_+=\sum_r\tilde\psi^\dag_{1,r} \tilde \psi_{1,r}$ and $Q_-=\sum_s\tilde\psi^\dag_{2,s} \tilde \psi_{2,s}$.
One can obtain \eqref{Q+} based on the zeta-function regularization:
\be
\sum_{r\in\mathbb Z +a,\,r<0}1=\zeta(0,1-(a-[a]))=a-[a]-1/2,
\ee
where $0<a<1$, and zeta-function is defined as $\zeta(s,x)=\sum_{n=0}^\infty (n+x)^{-s}$. Similarly, to regularize $Q_-$, one can use
\be
\sum_{s\in\mathbb Z -a,\,s<0}1=\zeta(0,1-(a-[a]))=-a-[a]+1/2,
\ee
based on which one can obtain 
\be
\label{Q-}
Q_-=\sum_{s\in \mathbb Z-\alpha/2\pi}: \tilde{\psi}_{1,r}^\dag \tilde{\psi}_{1,r}:
-\frac{\alpha}{2\pi}-\Big[\frac{\alpha}{2\pi}\Big]+\frac{1}{2},
\ee
where $0<\alpha<2\pi$. Then the 2-form $\omega^{(2)}$ in \eqref{F2} can be obtained by considering
\be
\small
\omega^{(2)}=\left(
\langle Q_+\rangle-\langle Q_-\rangle
\right)\Omega_+^{(2)}=\left(\frac{\alpha}{\pi}-1\right)\Omega_+^{(2)}.
\ee
Note that while $
\langle Q_+\rangle-\langle Q_-\rangle$ varies, 
 the total fermion number, $\langle Q_+\rangle + \langle Q_-\rangle$, remains unchanged during the pumping process.

\begin{figure}[tp]
\centering
\includegraphics[width=2.2in]{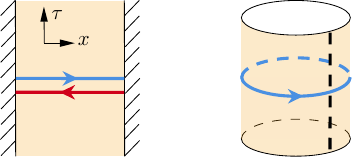}
\caption{Mapping from a BCFT with parametrized conformal boundary conditions to a chiral CFT on a cylinder with parametrized conformal interfaces/defects. 
The blue/red solid lines correspond to the right/left moving components.}
\label{BCFT_CFT}
\end{figure}

\section{From parametrized conformal boundaries to 
parametrized conformal interfaces}

For the parametrized BCFTs as studied in this work, they can be mapped to chiral CFTs on a circle with twisted boundary conditions, or more generally parametrized conformal interfaces.
It is emphasized that the resulting physics is fundamentally different. In particular, realizing the corresponding flow of charge or Berry curvature in the chiral theory may require coupling to a higher-dimensional bulk system, i.e., to cancel the anomaly in the parametrized chiral CFTs.

In this appendix we illustrate this mapping based on Dirac fermion CFTs.
Let us first illustrate with the single Dirac fermion as studied in the main text. As shown in Fig.\ref{BCFT_CFT}, the right-moving field and left-moving field are coupled at the two boundaries as $\psi_L(x)=\psi_R(x)$ at $x=0$, and $\psi_R(x)=e^{i\alpha} \psi_L(x)$ at $x=l$. We can map this BCFT to a \textit{chiral} CFT on a circle $[-l, l]$ where the two ends are identified, 
by defining
\be
\left\{
\begin{split}
&\psi_R'(x)=\psi_R(x), \quad \text{if }x\in[0,l]\\
&\psi_R'(x)=\psi_L(-x), \quad \text{if }x\in [-l,0].
\end{split}
\right.
\ee
After this ``unfolding'' procedure,  the parametrized boundary conditions are mapped to the following twisted boundary conditions for the chiral CFT:
\be
\psi_R'(x+L)=e^{i\alpha}\psi_R'(x), \quad L=2l.
\ee

Similarly, if we have two right-moving chiral Dirac fermions, $\psi_1$ and $\psi_2$, living on a circle of length $L=2l$, one can introduce the $\SU(2)$ twisted boundary conditions as 
\be
\Psi(x+L)=U\Psi(x), \quad U\in \SU(2),
\ee
where $\Psi(x)=(\psi_1(x),\psi_2(x))^T$.
This case corresponds to our example with Berry curvature flow.

\section{From conformal boundary conditions to 
conformal boundary states}

In this appendix, we demonstrate how a parametrized family of conformal boundary conditions leads to a corresponding family of conformal boundary states.
We follow the systematic approach for free field theories 
as introduced in \cite{hori2003mirror}.

\begin{figure}[tp]
\centering
\includegraphics[width=1.8in]{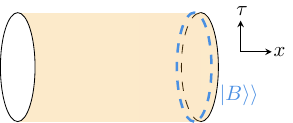}
\caption{A cylinder in $w=x+i\tau$ with $\tau=\tau+2\pi$, and an open boundary is imposed on the right end. By considering the Wick rotation $w'=e^{-i\pi/2}w$, one can define the conformal boundary state in the Hilbert space of the theory associated to the boundary (dashed circle).
}
\label{Boundary_State}
\end{figure}

We first rewrite the boundary conditions on a Euclidean Riemann surface $w=x+i\tau$ with boundary circles, as shown in Fig.\ref{Boundary_State}. That is, we compactify the Euclidean time direction as $\tau=\tau+2\pi$. 
Next, we perform a $\pi/2$ rotation of the coordinate as
\be
\label{w'}
w'=x'+i\tau'=e^{-i\pi/2}w,
\ee
or equivalently $(x',\tau')=(\tau,-x)$.
Then the new coordinates $x'$ and $\tau'$ can 
be considered as the new space and time.
 
With the Wick rotation in \eqref{w'}, for the fields $\psi_R$ and $\psi_L$, we have 
$\psi_R\sqrt{dw}=\psi_R\, e^{i\pi/4}\sqrt{dw'}$,
$\psi_R^\dag \sqrt{dw}=\psi_R^\dag \, e^{i\pi/4}\sqrt{dw'}$,
$\psi_L\sqrt{d\bar w}=\psi_L\, e^{-i\pi/4}\sqrt{d\bar {w'}}$,
and $\psi_L^\dag\sqrt{d\bar w}=\psi_L^\dag\, e^{-i\pi/4}\sqrt{d\bar {w'}}$, respectively. It is appropriate to introduce the notation
\be
\label{Rotate_field}
\psi_R'=e^{i\pi/4} \psi_R, \quad \psi_R'^\dag=e^{i\pi/4} \psi_R^\dag, 
\ee
and similarly 
$
\psi_L'=e^{-i\pi/4} \psi_L$ and $\psi_L'^\dag=e^{-i\pi/4} \psi_L^\dag$. One should be careful that in \eqref{Rotate_field}, the same phase factor $e^{i\pi/4}$ appears in both equations .

Now let us first consider the conformal boundary conditions in 
\eqref{U1_boundary}, i.e., $\psi_R=e^{i\alpha} \psi_L$.
With the notation in \eqref{Rotate_field}, the boundary condition becomes
\be
\label{BC_rotate}
\psi_R'=e^{i\pi/2} e^{i\alpha}\, \psi_L', \quad 
\psi_R'^\dag=e^{i\pi/2} e^{-i\alpha}\, \psi_L'^\dag.
\ee
On the $2\pi$ circle, we consider the mode expansion as follows:
\be
\label{Mode_expand}
\psi'_R=\sum_{r\in \ZZ+a}\psi_{R,r} \, e^{ir x'},
\quad
\psi'_L=\sum_{s\in \ZZ+\tilde a}\psi_{L,s} \, e^{-is x'}
\ee
where we have considered twisted boundary conditions specified by $a$ and $\tilde a$ along the circle. The ground state of the corresponding closed-string Hamiltonian defined on the $2\pi$ circle is denoted as $|0\rangle_{a,\tilde a}$, which is 
annihilated by $\psi_{R,r}(r\ge 0)$, $\psi^\dag_{R,r} (r<0)$, $\psi_{L,s}(s\ge 0)$, and $\psi^\dag_{L,s} (s<0)$. 
One can find that the boundary condition in \eqref{BC_rotate} is consistent with the mode expansion in \eqref{Mode_expand} only if $a=-\tilde a$ (mod $\ZZ$). Then, in terms of the modes $\psi_{R,r}$ and $\psi_{L,s}$, the conformal boundary conditions become
\be
\psi_{R,r}=ie^{i\alpha}\, \psi_{L,-r},\quad
\psi^\dag_{R,r}=ie^{-i\alpha}\, \psi_{L,-r}^\dag,
\ee
where $r\in \ZZ+a$.
The conformal boundary state $|B\rangle\rangle$ is then constructed as the solution to:
\be
\begin{split}
\left( \psi_{R,r} - i e^{i\alpha} \psi_{L,-r} \right) |B\rangle\rangle &= 0, \\
\left( \psi^\dag_{R,r} - i e^{-i\alpha} \psi^\dag_{L,-r} \right) |B\rangle\rangle &= 0.
\end{split}
\ee
One can find the one-parameter family of conformal boundary states as
\begin{equation}
\label{CBS_S1}
\small
|B\rangle\rangle = \exp\Big(
i e^{-i\alpha} \sum_{r \le 0} \psi_{R,r} \psi^\dag_{L,-r}
+ i e^{i\alpha} \sum_{r > 0} \psi^\dag_{R,r} \psi_{L,-r}
\Big) |0\rangle_{a, -a}.
\end{equation}

Next, for the conformal boundary conditions in Eq.\eqref{S3_boundary}, the corresponding conformal boundary states can be constructed in a similar manner. By applying a unitary transformation in Eq.\eqref{U_transform}, the boundary conditions are brought into the diagonal form of Eq.\eqref{S3_bdry_diagonal}, allowing the boundary states to be obtained using the same procedure as above. Finally, by inverting the unitary transformation in Eq.\eqref{U_transform}, one can explicitly express the resulting three-parameter family of conformal boundary states in terms of the original fermion fields.

\section{Entanglement Hamiltonians in families of gapped ground states}

\begin{figure}[tp]
\centering
\includegraphics[width=2.2in]{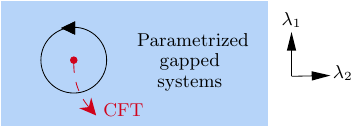}
\caption{Phase diagram in the parameter space spanned by $\lambda_1$ and $\lambda_2$.
We consider a closed surface (which is a circle here) surrounding the gapless point which is a CFT.
The parametrized family of gapped systems can be obtained by adding relevant perturbations to the CFT.
}
\label{Phase_diagram}
\end{figure}

In this appendix, we present a complementary approach to studying the topological properties of a family of gapped ground states near a critical point.

The motivation stems from the observation in \cite{Hsin_2020} that nontrivial higher Berry classes impose topological constraints on the many-body phase diagram, necessitating the existence of gapless points that are topologically protected. When these gapless points correspond to CFTs, the continuous family of gapped systems can be realized by introducing a continuous family of relevant perturbations to the CFT, as illustrated in Fig.\ref{Phase_diagram}. Our focus here is on such families of gapped systems in the vicinity of the CFT.

Let us consider a parametrized family of one dimensional gapped systems defined on $(-\infty,+\infty)$. 
When this family is nontrivial, there will be a flow of 
generalized charge, such as the $U(1)$ charge or Chern number, along this one dimensional system.
One should be able to see such flows by simply looking at the half space $A=[0,+\infty)$. All the information of subsystem $A$ is contained in the reduced density matrix
\begin{equation}
\label{rho_A}
\rho_A(\lambda) = e^{-2\pi \, H_E(\lambda)},
\end{equation}
which depend on parameters $\lambda \in X$.

As shown in \cite{Cho_2017_A, Huang_2024}, by analyzing the effects of relevant perturbations (in a setup distinct from that in Fig.~\ref{Fig:rho} of the main text), the entanglement Hamiltonian 
$H_E(\lambda)$ of a gapped system corresponds to the physical Hamiltonian of a BCFT defined on a finite interval of length $l=\log(\xi/a)$, where $\xi$ is the correlation length and $a$ is a UV cutoff. 
The two boundary conditions of this BCFT are as follows: one is a free boundary condition, representing the entanglement cut, which remains unaffected by the relevant perturbation; the other corresponds to an interface between the CFT and a gapped system, and is determined by the relevant perturbation.

Generalizing the above picture to our setting, the parametrized entanglement Hamiltonians $H_E(\lambda)$ in Eq.\eqref{rho_A} map to the \textit{physical} Hamiltonians of the BCFTs illustrated in Fig.\ref{Fig:BCFT}, where one boundary condition is fixed while the other is continuously parametrized by  $\lambda$, with the system size $\log(\xi/a)$.
Our results on multi-parameter spectral flow can then be directly applied to $H_E(\lambda)$, revealing its universal topological features. 
A detailed analysis using lattice models will be presented in a future work.

It is useful to remark on the difference and relationship between this appendix and the setup in Fig.\ref{Fig:rho}. In the regularized conformal boundary states of Fig.\ref{Fig:rho}, we begin with conformal boundary states and introduce irrelevant perturbations in the form of regularization factors. In contrast, the approach developed in \cite{Cho_2017_A} starts from a CFT and introduces a relevant perturbation. Despite these differences, both setups lead to the same physical outcomes, as the resulting states are connected to the ground states of gapped Hamiltonians.

\end{document}